\documentclass{PoS}
\usepackage{epsfig,multicol}

\newcommand{\be}{\begin{equation}}
\newcommand{\ee}{\end{equation}}
\newcommand{\beq}{\begin{eqnarray}}
\newcommand{\eeq}{\end{eqnarray}}

\newcommand{\s}{\widetilde{\sigma}}

\newcommand{\LMS}{\Lambda_{\overline{MS}}}
\newcommand{\mod}{{\rm mod}}
\newcommand{\downrightarrow}{\searrow}

\PoS{PoS(LAT2005)323}

\title{'t Hooft loops and perturbation theory}

\ShortTitle{'t Hooft loops and perturbation theory}

\author{\speaker{Philippe de Forcrand}\\
  Institute for Theoretical Physics, ETH Z\"urich,
  CH-8093 Z\"urich, Switzerland\\
  CERN, Physics Department, TH Unit.
  CH-1211 Gen\`{e}ve 23, Switzerland\\
  E-mail: \email{forcrand@phys.ethz.ch}}

\author{Biagio Lucini\\
  Institute for Theoretical Physics, ETH Z\"urich,
  CH-8093 Z\"urich, Switzerland\\
  E-mail: \email{lucini@phys.ethz.ch}}

\author{David Noth\\
  Paul Scherrer Institut,
  CH-5232 Villigen PSI, Switzerland\\
  E-mail: \email{nothd@phys.ethz.ch}}

\abstract{
We show that high-temperature perturbation theory describes
extremely well the area law of $SU(N)$ spatial 't Hooft loops,
or equivalently  the tension of the interface between different
$Z_N$ vacua in the deconfined phase. 
For $SU(2)$, the disagreement between Monte Carlo data
and lattice perturbation theory
for $\tilde\sigma(T)/T^2$ is less than 2\%, down to temperatures
${\cal O}(10)~T_c$. For $SU(N), N>3$, the ratios of interface
tensions, $(\tilde\sigma_k/\tilde\sigma_1)(T)$, agree with 
perturbation theory, which predicts tiny deviations from
the ratio of Casimirs, down to nearly $T_c$. 
In contrast, individual tensions differ markedly from the
perturbative expression.
In all cases, the required precision Monte Carlo measurements are made possible 
by a simple but powerful modification of the 'snake' algorithm.
}

\FullConference{XXIIIrd International Symposium on Lattice Field Theory\\
                 25-30 July 2005\\
                 Trinity College, Dublin, Ireland}

\begin{document}

\section{Introduction}
\label{sect:introduction}
\FIGURE[t]{
\epsfig{file=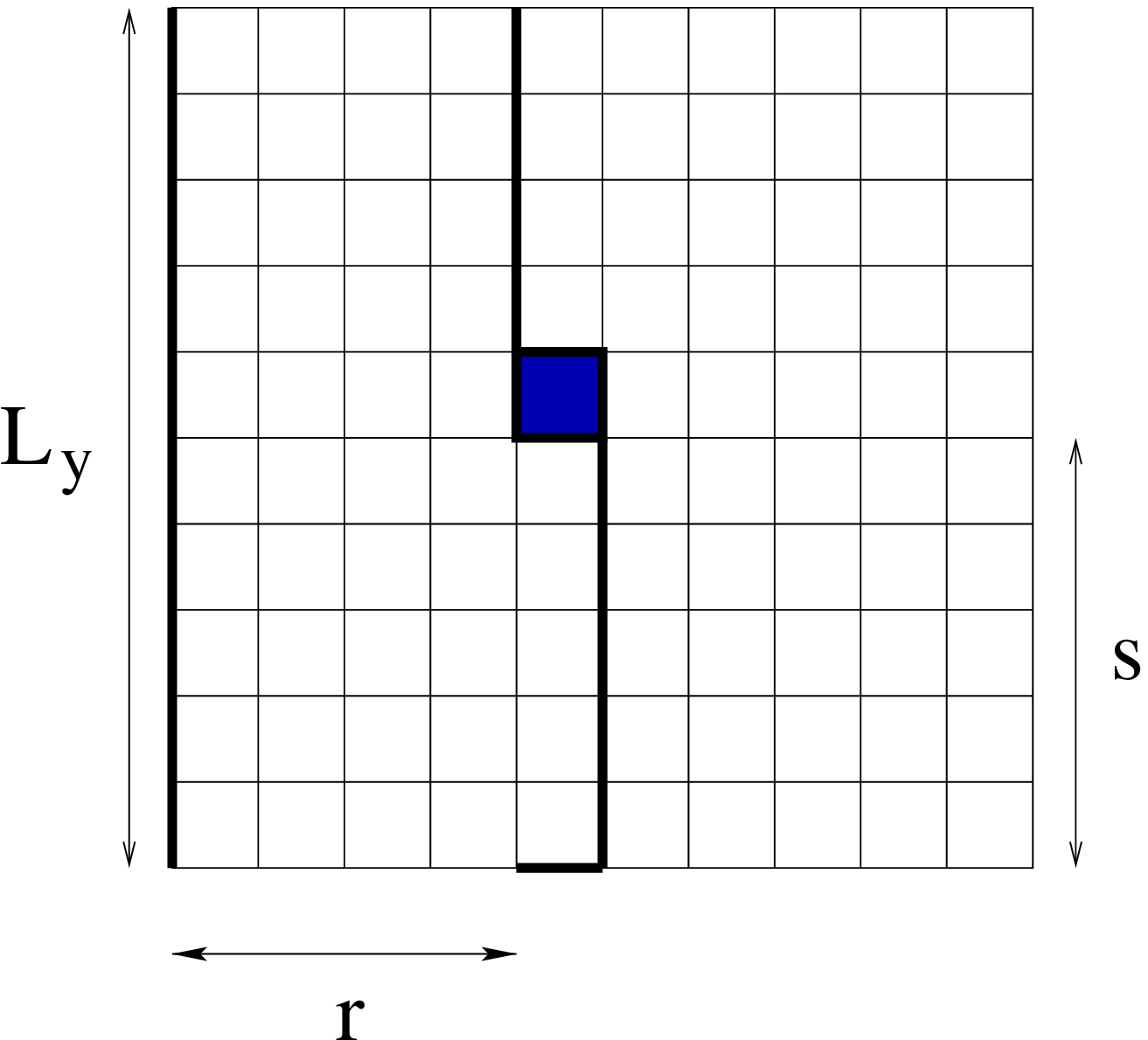,scale=0.5}
\caption{The interface tension is extracted from the measured ratio $\exp(-\s a^2)$
of partition functions of two systems with partial interfaces, which differ by the single shaded plaquette.}
\label{fig:1}
}
't Hooft has shown that, in Yang-Mills theories at zero temperature, and in the absence of massless
modes, an area law for the Wilson loop implies a perimeter law for the 't Hooft loop, and vice versa.
This dual behaviour carries over at finite temperature~\cite{PRL}: while the temporal Wilson loop
adopts a perimeter law in the deconfined phase $T > T_c$, the spatial 't Hooft loop acquires an area law.

This can be shown without any simulations. A spatial 't Hooft loop $\tilde W(\partial\tilde\Sigma)$
bounding a surface $\tilde\Sigma$, say, in the $(x,y)$ plane, is created on the lattice by ``flipping'' a stack of
$(z,t)$ plaquettes, one per plane pierced by $\tilde\Sigma$. ``Flipping'' means that the plaquette
matrix $U_P$ is multiplied by a center element $z_k = \exp(i\frac{2\pi k}{N})~{\bf 1}$ before its
trace is evaluated. The corresponding partition function $Z_{\rm flipped}$ gives the 't Hooft loop
expectation value via $\langle \tilde W(\tilde\Sigma) \rangle = Z_{\rm flipped}/Z_{\rm pbc}$,
where the denominator corresponds to the usual action, with periodic boundary conditions.
 
It is easy to move the stack of plaquettes to a corner of the $(z,t)$ plane, then absorb the phase
factor $z_k$ in the boundary condition of the temporal link $U_t$: 
\be
U_t(z + L_z,t=t_0) = z_k U_t(z,t=t_0)
\ee
$Z_{\rm flipped}$ then becomes the usual partition function (no flipped plaquettes) of a system
where a twist has been enforced on the Polyakov loop $P$:
\be
P(x,y,z + L_z) = z_k P(x,y,z) ~~ {\rm for} ~ x,y \in \tilde\Sigma
\ee
When $T > T_c$, the center symmetry is broken, and the twist above causes an interface to appear,
perpendicular to the $z$-direction, because the Polyakov loop lies in different $Z_N$ sectors
at $z=0$ and $z=L_z$. The associated increase in free energy can be ascribed to
the interface tension, or equivalently to the 't Hooft loop [dual] string tension $\s$. These
are two names for the same observable.

This equivalence allows us to compare numerical results for $\s$ with old perturbative calculations
of the interface tension~\cite{Bhat}. It also suggests a simple way to measure $\s$,
further simplifying the ``snake'' algorithm~\cite{LAT04}:
just increase the interface area by one plaquette, and measure the change in free energy
$\exp(-\s a^2)$~(see Fig.~1). 
We perform such simulations and comparisons for $SU(N)$, discussing first the case $N=2$~\cite{Noth}
and then $N\ge 3$.

\section{SU(2)}
\label{sect:su2}

We have repeated the 1-loop perturbative calculation of the interface tension of \cite{Bhat},
on a lattice of $N_t$ time-slices. In fact, it amounts to substituting 
$p \rightarrow \hat p = \frac{2}{a} \sin\frac{p a}{2}$
in the fluctuation determinant. The resulting interface tension is multiplied by a coefficient
$C_{\rm lat}(N_t)$, whose difference with 1 is shown in Fig.~2. 
The expected $1/N_t^2$ behaviour does not set in until
$N_t \gtrsim 10$, and for practical values of $N_t$, $C_{\rm lat}(N_t)$ is large and not
even monotonic. Thus, it provides an essential correction when computing $\s$ on the lattice.
The corresponding $SU(N)$ leading-order perturbative prediction is
\be
\frac{\tilde\sigma_k}{T^2}(T) = 
C_{\rm lat}(N_t) \times \frac{4 \pi^2}{3\sqrt{3}} \times 
\frac{1}{\sqrt{(g^2 N)(T)}} \times k(N-k)
\ee
Using the ``snake''-like algorithm above, we have measured $\s$ in $SU(2)$ for
a wide range of couplings $\beta$ and lattice sizes $N_t$. As shown 
Fig.~3, excellent agreement is found with perturbation \nolinebreak theory at 
${\cal O}(g^2)$~\cite{Bhat}, after converting the coupling from the improved 
scheme~\cite{Huang} to the lattice for $SU(2)$:
\be
\frac{\s}{T^2}(\beta,N_t) = C_{\rm lat}(N_t) \times \frac{2 \pi^2}{3\sqrt{6}} 
\times \sqrt{\beta} \times (1 - (0.21467 + 0.04644 \log N_t) \frac{4}{\beta})
\label{SU2}
\ee
The unknown ${\cal O}(g^2)$ contribution to $C_{\rm lat}(N_t)$ appears 
to be very small. We can then convert $(\beta,N_t)$ to temperatures
$T/\LMS$, using 2-loop perturbation theory and $T_c/\LMS = 1.31(8)$~\cite{fingberg}.
The data obtained at various lattice spacings collapse nicely, and the
ratio of the measured over the perturbative $\s$ remains within $\sim 2\%$
of 1, down to ${\cal O}(10) T_c$ where the measured $\s$ decreases, 
since it has to vanish at the second-order transition $T=T_c$. See Fig.~4.

\begin{figure}[t]
\vspace*{-0.3cm}
 \begin{minipage}[t]{0.45\textwidth}
 \epsfxsize=1.0\textwidth {\epsfbox{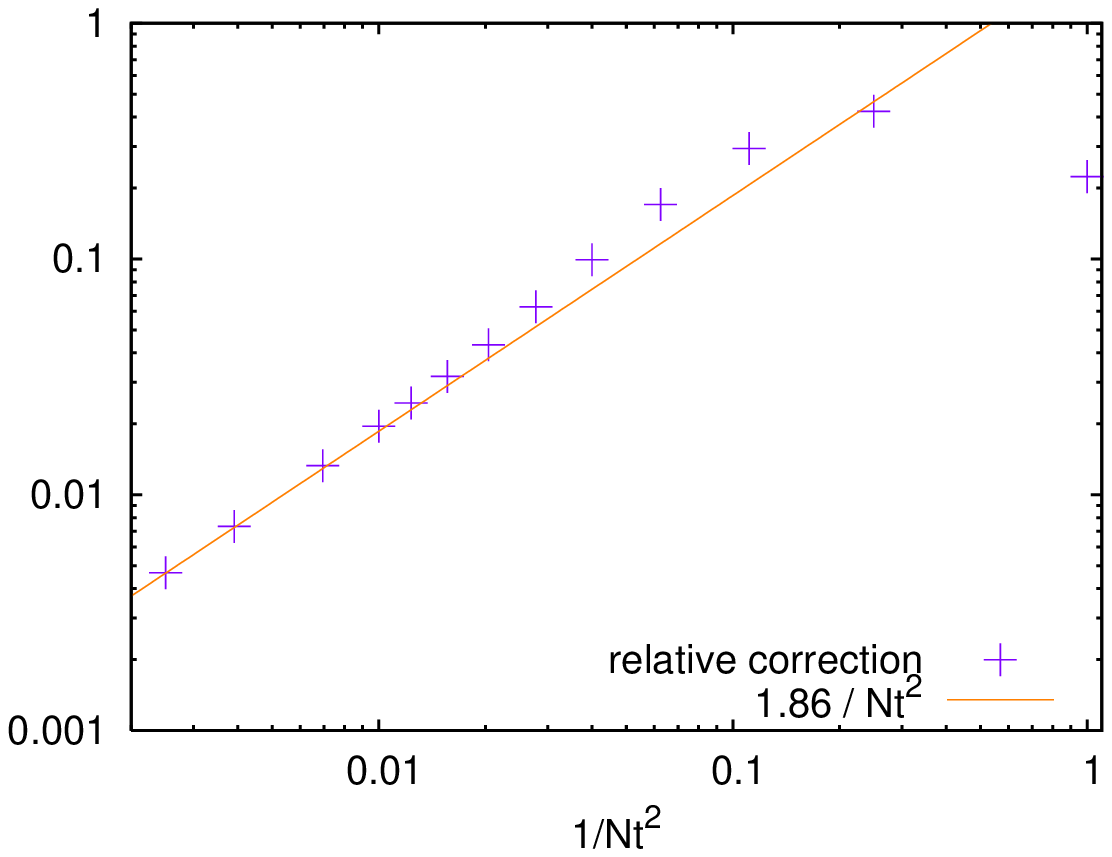}}
 \caption[Figure 2]
{The relative correction $(C_{\rm lat}(N_t) -1)$ to the interface tension $\s$,
for different temporal lattice sizes $N_t$.}
 \end{minipage}
 \hfill
 \begin{minipage}[t]{0.45\textwidth}
 \epsfxsize=1.0\textwidth {\epsfbox{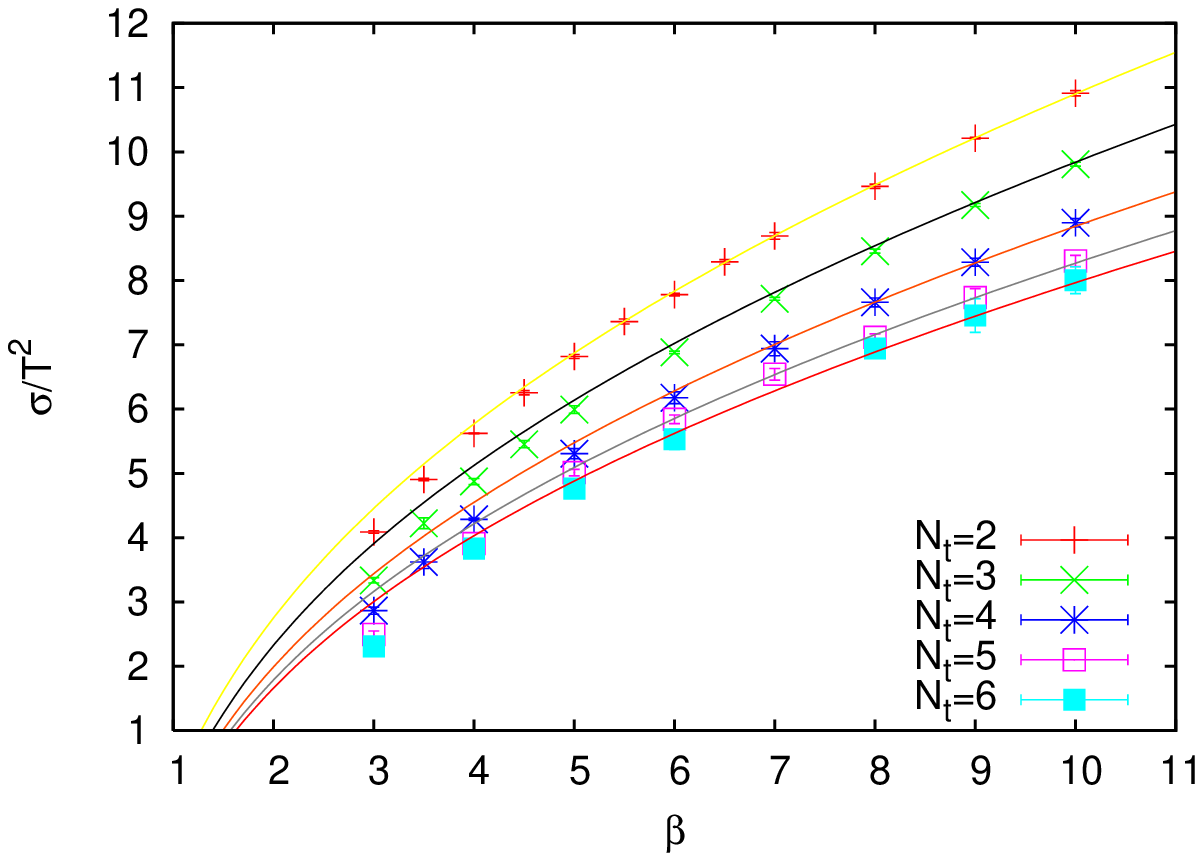}}
 \caption[Figure 3]
{The dimensionless ratio $\s/T^2(\beta,N_t)$ agrees excellently with the
parameter-free perturbative formula eq.(\ref{SU2}).}
 \end{minipage}
\vspace*{-0.1cm}
\end{figure}

\begin{figure}[bh!]
\vspace*{-0.2cm}
\centerline{\epsfig{file=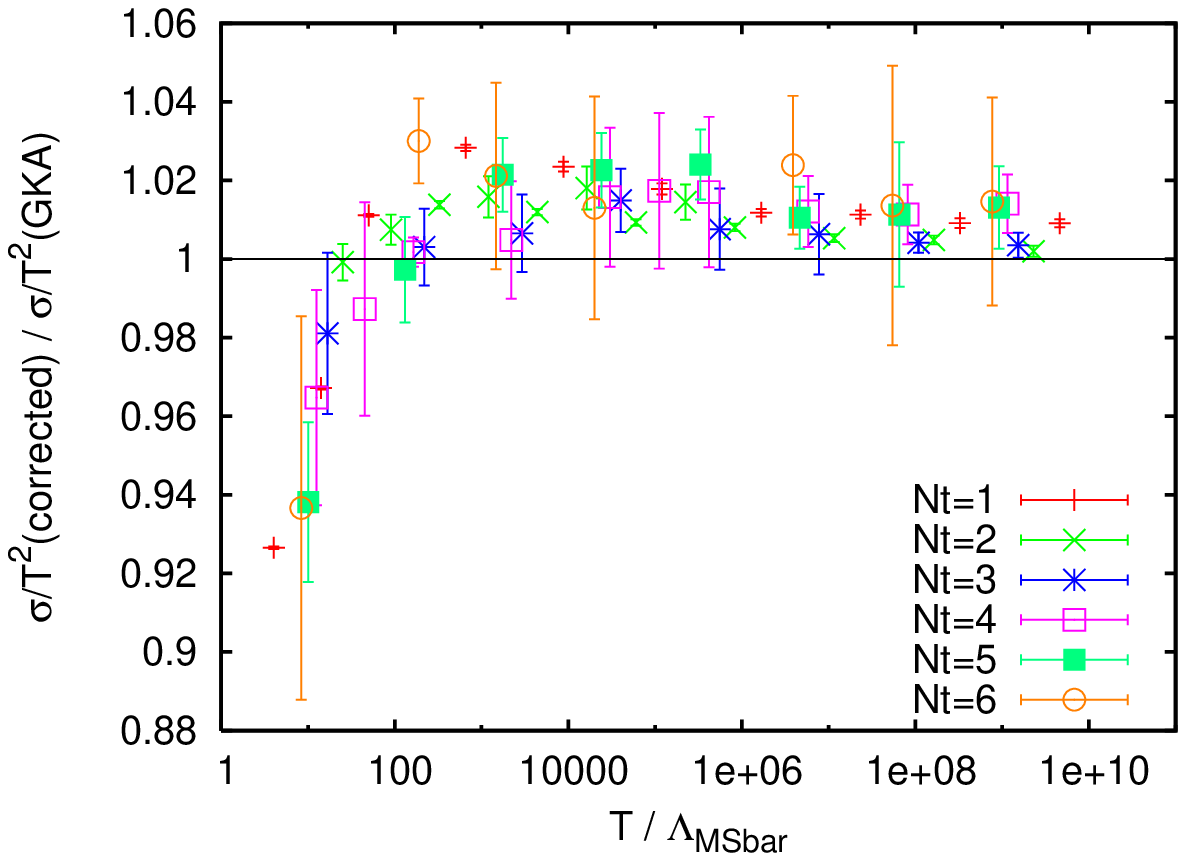,scale=0.53}}
\vspace*{-0.2cm}
\caption{The ratio of the measured over the perturbative interface tension,
as a function of temperature. Deviations are $\lesssim 2\%$, down to the vicinity 
of $T_c$ where the measured $\s$ must vanish.}
\end{figure}

\FIGURE[t]{
\epsfxsize=1.0\textwidth {\epsfbox{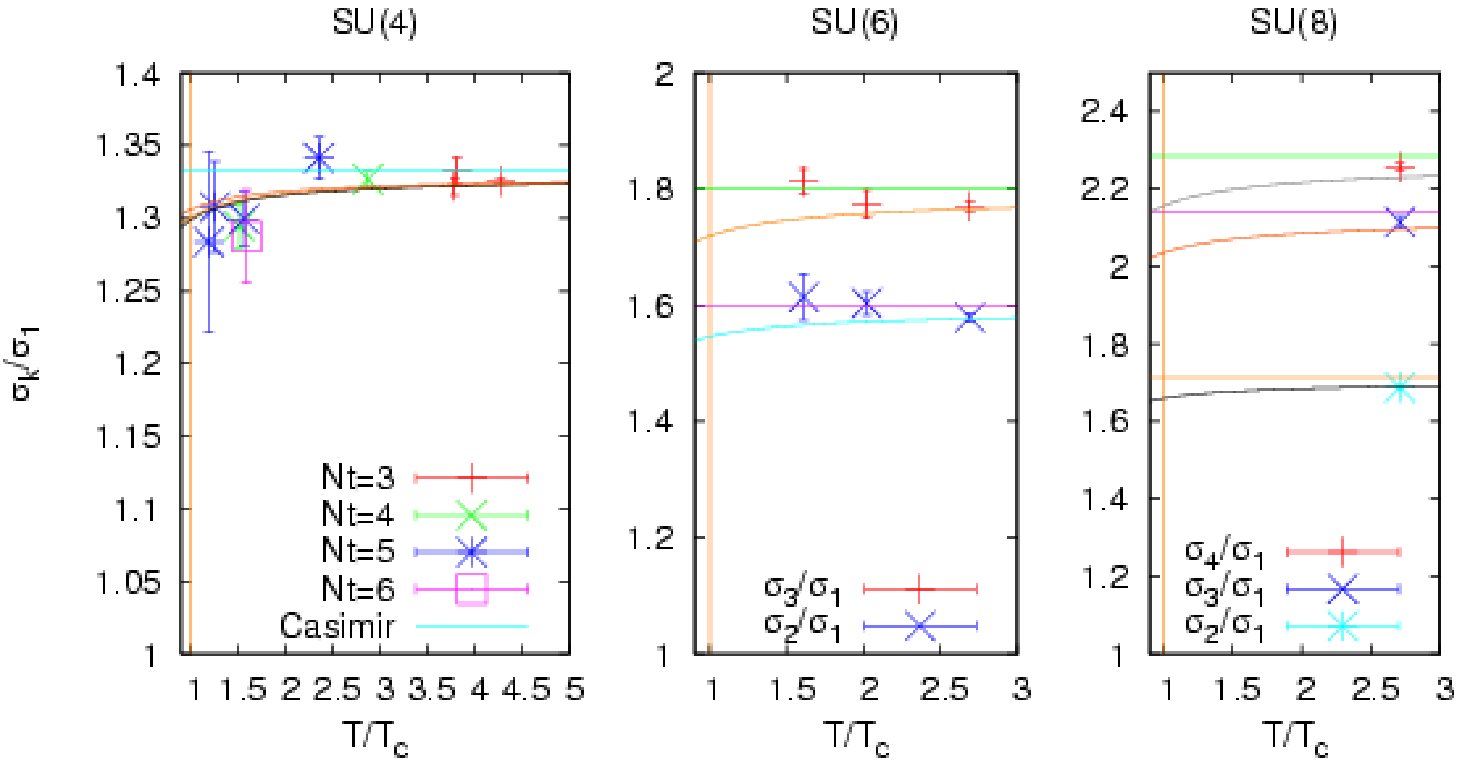}}
\caption{Ratios of interface tensions $\s_k/\s_1$, as a function of temperature,
for $SU(4)$ ({\em left}), $SU(6)$ ({\em middle}) and $SU(8)$ ({\em right}).
The horizontal lines mark the Casimir values $\frac{k(N-k)}{N-1}$. 
The curves show the ${\cal O}(g^3)$ perturbative prediction of~\cite{CPKA}.
The 2 curves for $SU(4)$ correspond to $T_c/\LMS = 1.10$ and $1.35$.}
}

\DOUBLEFIGURE[t]
{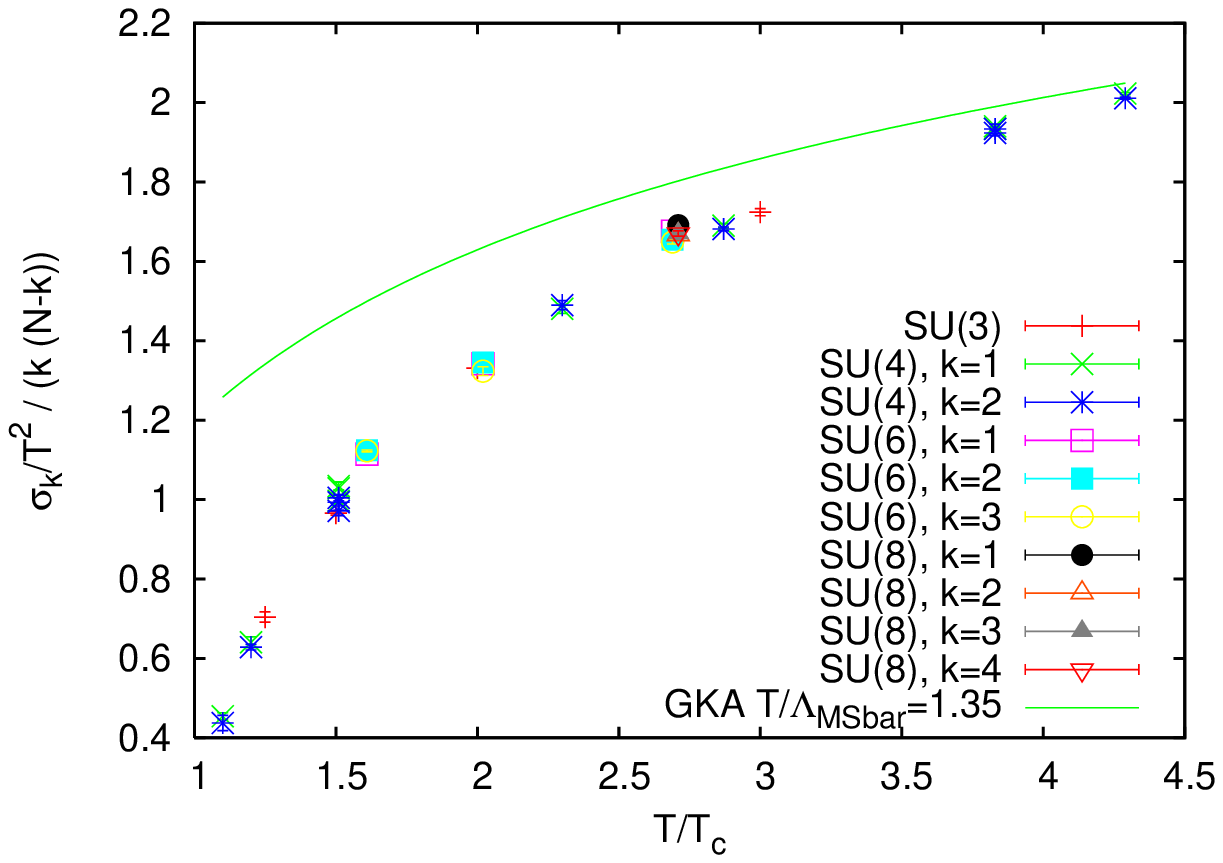,scale=0.55,angle=0}{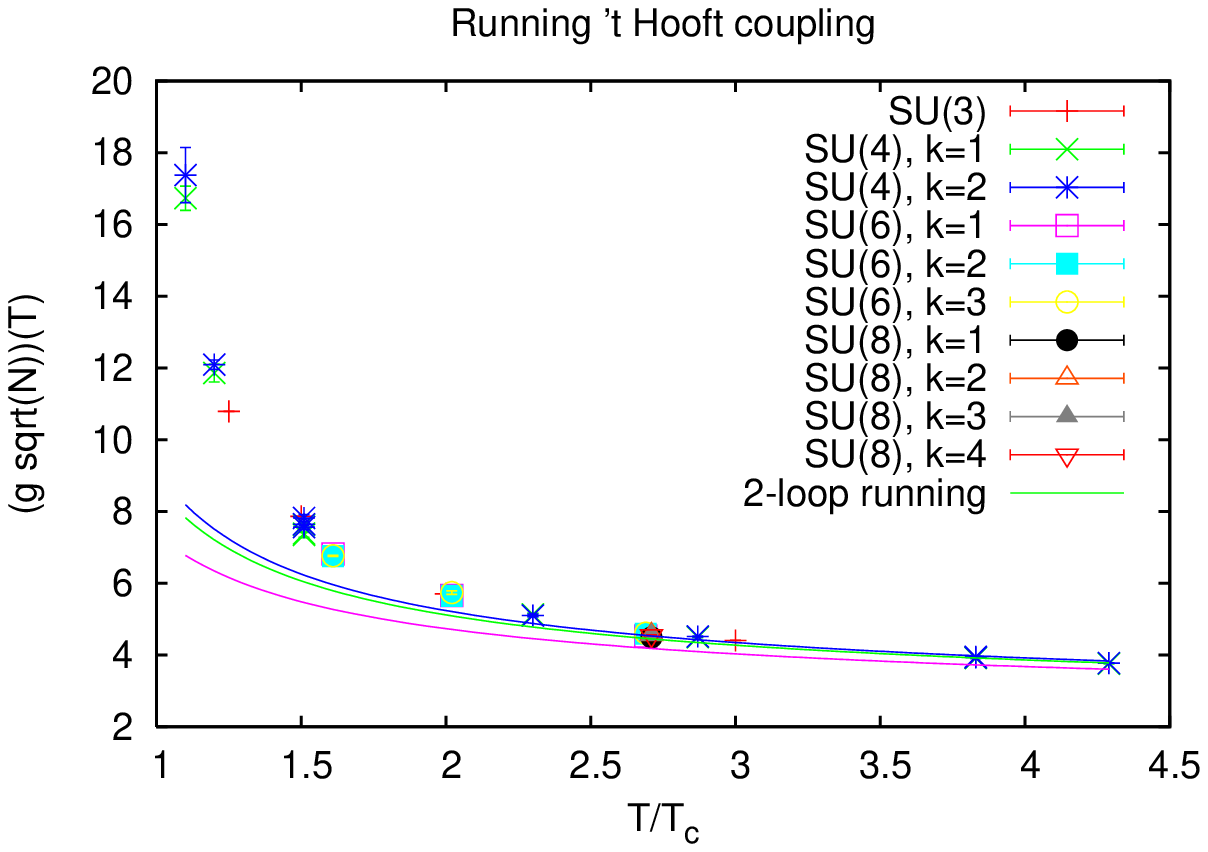,scale=0.55,angle=0}
{$\s_k/T^2$, divided by the Casimir $k(N-k)$, as a function
of temperature. Data for all $SU(N), N\geq 3$ gauge groups and all values 
of $k$ collapse on a single curve, which deviates considerably from the
perturbative prediction (solid curve) which worked so well for $\s_k/\s_1$~Fig.~5.
}
{Running 't Hooft coupling $(\tilde g \sqrt{N})(T)$ extracted from $\s_k/T^2$, 
as a function of temperature.
The solid lines show 2-loop running, with $T_c/\LMS=1.10, 1.25$ and $1.35$.
}

\section{SU(N)}
\label{SUN}

The same numerical method can be used for $SU(N)$ gauge theories. 
The interesting difference is that the $N$-fold degeneracy of the vacuum now allows
for inequivalent interfaces. In the vacuum, the Polyakov loop can take values 
$z_k = \exp(i 2\pi \frac{k}{N}) ~ {\bf 1}$~$\in Z_N$.
Interfaces separating two vacua $k_1, k_2$ rotate the Polyakov loop by 
$z_{k_1} z_{k_2}^{-1} = \exp(i 2\pi \frac{k_1-k_2}{N})$.
The corresponding interface tension is $\s_k$, with $k=k_1-k_2 ~\mod ~N$.
Charge conjugation imposes $\s_k = \s_{N-k}$.
This leaves ${\rm int}(\frac{N}{2})$ independent interface tensions.

We measure them independently using the same setup of Fig.~1, where the 
``twisted'' plaquette has action ${\rm ReTr} ~z_k U_P$.
The temperature can be varied by changing the inverse coupling $\beta$ or the number
$N_t$ of time-slices. To stay clear of a bulk first-order transition for $N>3$,
we chose $N_t \geq 5$ for $T/T_c \le 1.5$. The finite-temperature transition is first-order, and its
strength increases with $N$. This allows us to consider spatial sizes 
$N_s = 3 N_t$ only, and still maintain good control over the thermodynamic
limit. We check this by varying $N_s$ in the potentially most problematic
regime, for $SU(4)$ near $T_c$: no measurable finite-size effect is found.

Like for $SU(2)$, we can compare our numerical results with perturbation theory, 
which is available to ${\cal O}(g^3)$~\cite{Bhat}:
\be
\frac{\tilde\sigma_k}{T^2}(T) = 
\frac{4 \pi^2}{3\sqrt{3}} \times 
\frac{1}{\sqrt{(g^2 N)(T)}} \times k(N-k)
\times (1 - 15.27853.. ~ \frac{(g^2 N)(T)}{16 \pi^2})
\ee
Systematic errors of any kind will tend to cancel in the ratio $\s_k/\s_1$.
Our results are presented in Fig.~5 as a function of $T/T_c$, for gauge groups
$SU(4), SU(6)$ and $SU(8)$~\footnote{The temperature is obtained from the
determination of the $T=0$ string tension at the same coupling $\beta$~\cite{Lucini}.}. 
The accuracy on $\s_k/\s_1$ is 1-3\%.
In the $SU(4)$ case, a scaling test (from $N_t=5$ to 6 time-slices) shows no
significant scaling violations. The horizontal lines in Fig.~5 correspond to 
the leading and subleading order perturbation theory, i.e. the ratio of Casimirs
$\frac{k(N-k)}{N-1}$.
One can see excellent agreement with the data at high temperature, as expected,
with only tiny downward deviations at lower temperatures $T \downrightarrow T_c$.

Moreover, these tiny deviations are consistent with the bending curves in Fig.~5,
which show the ${\cal O}(g^3)$ perturbative prediction~\cite{CPKA}. The latter is
expressed as a function of $T/\LMS$, whereas in our simulations we fix $T/T_c$.
The needed factor $T_c/\LMS$ is not known accurately, but a large variation between
the accepted bounds 1.10 - 1.35 \cite{Laine} in $SU(4)$ has almost no visible effect:
the corresponding curves in Fig.~5(left) are almost indistinguishable.
Thus, agreement of $\s_k/\s_1$ with ${\cal O}(g^3)$ perturbation theory persists in 
all our simulations at all temperatures studied. Disagreement, if it occurs, should be 
most visible for $SU(4)$, since this is the case where the deconfinement transition
is the weakest. This is also the most numerically difficult case, because the 
interface tension becomes quite small (thus harder to measure) as $T\downrightarrow T_c$.
Nevertheless, we observe consistency with perturbation theory, even for the smallest
temperature where we can maintain sufficient accuracy, namely $T/T_c \sim 1.1$.

Since agreement with perturbation theory is so good for $\s_k/\s_1$, one might expect
the same for individual $k$-tensions $\s_k$. This is not at all the case, as shown
in Fig.~6. This figure shows $\frac{\s_k/T^2}{k(N-k)}$, as a function of $T/T_c$,
for all the gauge groups $SU(N), N>2$ and $k$-values we have considered. 
The solid line in the figure shows the ${\cal O}(g^2)$ perturbative prediction,
which is independent of $N$ and $k$ (The $N$-, $k$-dependent ${\cal O}(g^3)$ correction 
is very small). Large deviations from perturbation theory are visible, showing that
the interface tension sharply decreases near $T_c$, a phenomenon not captured by
the perturbative expansion which is blind to the phase transition.
Nevertheless, the departure from perturbation theory appears to be universal:
data for all gauge groups collapse, even for $SU(3)$ (The $SU(2)$ data, which are not shown,
lie significantly below the rest).
Thus, the $SU(N=\infty)$ limit is approached very fast, and one single curve
$\frac{\s_k/T^2}{k(N-k)}(\frac{T}{T_c})$ 
appears sufficient to give a good, non-perturbative description of all $SU(N),N>2$
interface tensions at all temperatures.
This universal, non-perturbative dependence can be expressed via a running 
coupling constant, whose definition is taken to agree with the lowest order 
perturbative prediction:
\be
(\tilde g\sqrt{N})(T) ~~~{\rm defined~by}~~~
\frac{\tilde\sigma_k}{T^2}(T) = \frac{4 \pi^2}{3\sqrt{3}} \times 
\frac{1}{\sqrt{(\tilde g^2 N)(T)}} \times k(N-k)
\ee
This running coupling is shown in Fig.~6 as a function of $T/T_c$, together
with the 2-loop running coupling, where the curves correspond to 
$T_c/\LMS = 1.10$, 1.25 and 1.35. Agreement occurs for $T \gtrsim 3 T_c$.
At lower temperatures, the coupling defined by the interface tension
rises faster than perturbation theory would predict. A similar phenomenon 
can be seen in other quantities, e.g. the $SU(N)$ pressure~\cite{pressure}.
Therefore, one might try to explain the two together, using the single
coupling $(\tilde g\sqrt{N})(T)$ and low-order perturbation theory.
This naive attempt fails quantitatively. Therefore, the pressure and the
interface tension give us two independent properties of the $SU(N=\infty)$
theory, which are related by non-perturbative dynamics.

Note that Ref.~\cite{Bursa} has performed a similar $SU(N)$ study to ours,
but has measured instead derivatives of $\s_k$ with respect to the
lattice coupling, $d\s_k/d\beta$. While their findings for this quantity,
i.e. Casimir scaling,
agree with ours for the $\s_k$'s, the claim they make (Casimir scaling
of the $\s_k$'s down to $T \sim 1.02 T_c$) is not substantiated by their data. 
Casimir scaling of
the $\s_k$'s implies the same for the derivatives, but the converse is
not true, because the integration constant (which dominates the value
of $\s_k$ near $T_c$) is not determined in \cite{Bursa}.
Our findings provide a justification for their claim.

\bibliographystyle{JHEP}

\end{document}